# LOW-VELOCITY IMPACTS ON TARGETS CONTAINING EMBEDDED CARBON NANOTUBES.

J. A. Carmona, M. Cook, J. Schmoke, J. Reay, and T. W. Hyde, CASPER (Center for Astrophysics, Space Physics and Engineering Research), Baylor University, Waco, TX 76798-7310 E-mail: Truell_Hyde@baylor.edu.

**Introduction:** A one stage Light Gas Gun (LGG) at CASPER [1] was employed to test the shielding capabilities of tiles composed of four different laminated nanotube combinations. These target tiles were named CSNEAT1, CSCNT1, HYCNTUT1 and HYCNTT1. For calibration purposes, a 3003-aluminum plate was also impacted and the craters formed on the various composition tiles compared.

**Procedure:** The CASPER LGG uses Helium to propel various size impactors down an instrumented beam line. For this experiment, 3/32" aluminum and chromed steel particles were employed. Target tiles were clamped uniformly at each corner (measured at 9 lb/in$^2$) with a second impact (witness) plate placed 10 cm behind the target. Both laser fan detectors and PZTs were employed to measure impactor velocities before and after impact. Two laser fans located within the beamline measure the speed of the projectile immediately after it leaves the barrel of the gun with these signals sent to a Tektronix 3032 oscilloscope. The frame holding the target tile is also instrumented via a PZT (donated by Dr. Harry D. Shirey, Piezo-Kinetics) installed within the frame. Thus at the time of impact, the signal from the second laser fan can be compared to that from the PZT supplying a second coincident measurement for the speed of the projectile. The same frame holding the target tiles also secures the second witness plate mentioned above. A second PZT located on this frame is used to detect impactor penetration of the target tile (if any) as well as measure the speed of the particle after penetration.

**Results:** LabVIEW was used to collect and record all experimental data with the results for 3/32" aluminum projectiles shown in Figures 1 through 4. Corresponding data obtained for 3/32" chromed steel projectiles is shown in Figures 5 through 8.

Fig. 1. Shallower crater depths and larger crater diameters than those resulting for identical impacts on the aluminum plate are apparent.

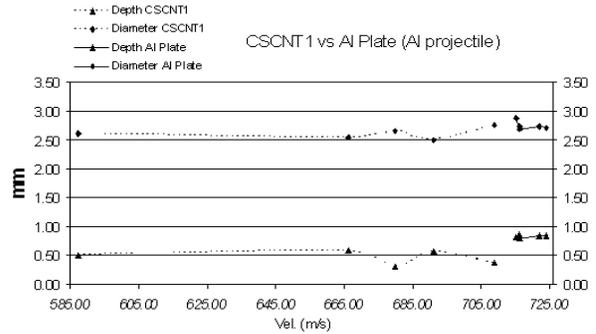

Fig. 2. The target tile shows crater depths shallower than those produced on the aluminum plate, but crater diameters closer to the crater diameters produced on the aluminum plate.

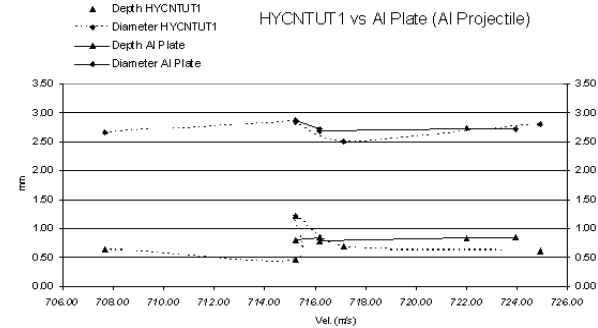

Fig. 3. For this sample, impactor damage is comparable to that seen on the aluminum plate. The trend and magnitude of the resulting crater damage is also comparable between target tile and aluminum control.

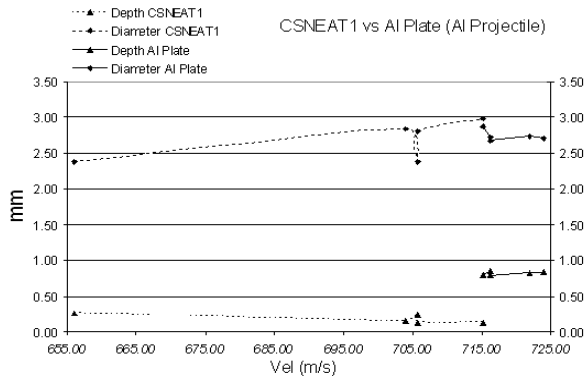

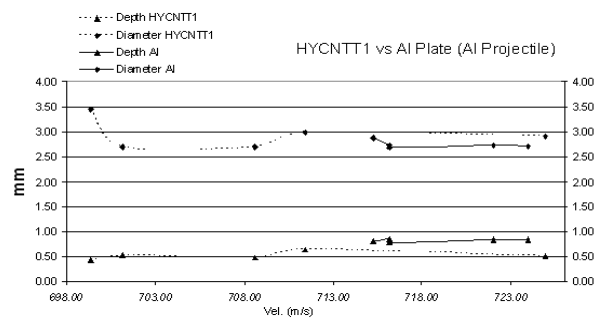

Fig. 4. Crater depth is shallower on the target tile than on the aluminum plate and crater diameters are slightly larger than the ones produced on the aluminum plate.

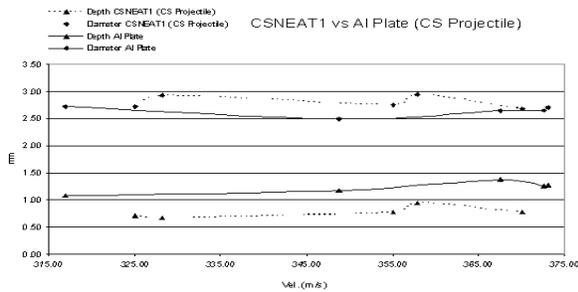

Fig. 5. In this sample (3/32" chromed steel impactors) the crater depth produced in the target tiles is shallower than the corresponding crater depth in the aluminum plate.

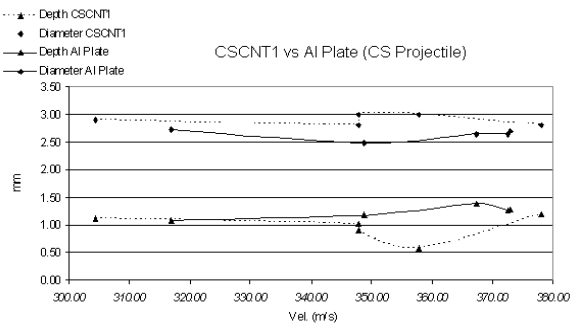

Fig. 6. Again, crater depth produced in the target tiles is shallower than the corresponding crater depth in the aluminum plate.

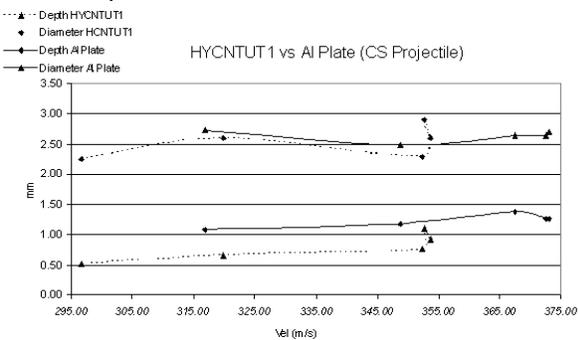

Fig. 7. A sudden increase in both crater depth and diameter can be seen at an impactor speed of 355 m/s. This result was only observed for this target tile

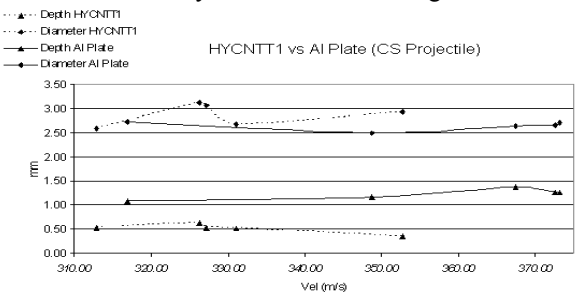

Fig. 8. Crater depth for the target tiles is smaller than those produced on the aluminum plate while crater diameter for both the aluminum plate and the target tiles steadily increases with increasing velocity.

In Figures 1-4, the impacting projectile was fired using a LGG pressure of 1500 psi, delivering an average energy to the target of approximately 4.5 Joules at an average impactor speed of 700 m/s. In Figures 5-8, the impacting projectile was fired using a pressure of 1500 psi, delivering an average energy of 3.0 Joules to the target at an average impactor speed of 340 m/s.

**Conclusions:** Figures 9 and 10 show the mechanical response for a representative target and 3003-aluminum plate for impacts taken under identical conditions. As seen, the nanotube targets resist impacts in a different manner than do targets composed of standard materials. Unfortunately, the data collected within this study does not provide adequate statistical information to determine either a clear trend or identify overall material behavior. As such, more data is necessary before a conclusion can be made whether this new material can be successfully employed as an appropriate shielding mechanism.

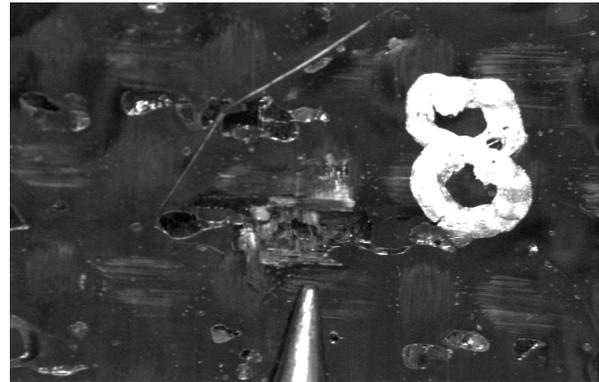

Fig. 9. The damage on the target does not leave a distinct crater; however, delamination of the tile due to its underlying carbon nanotube structure is apparent.

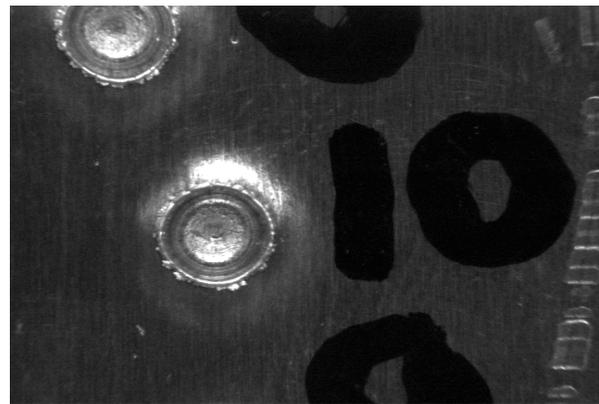

Fig. 10. Typical crater after impact on the 0.5" thick 3003-aluminum plate

**References:** [1] Carmona, et al. (2004) *LPS XXXV*, Abstract **#**1019.